\documentclass[twocolumn]{aastex631}
	\graphicspath{{./}{figures/}}

	\shorttitle{Characterizing Vortex-Driven Dynamics in the Solar Atmosphere}
	\shortauthors{Silva et al.}
	
\begin{document}
			
			\title{Characterizing Vortex-Driven Dynamics in the Solar Atmosphere Using Information Theory}

			\correspondingauthor{Suzana S. A. Silva et al.}
			\email{suzana.silva@sheffield.ac.uk}
			
			\author[0000-0001-5414-0197]{Suzana S. A. Silva}
			\affiliation{Plasma Dynamics Group, School of Electrical and Electronic Engineering, University of Sheffield, Sheffield, S1 3JD, UK
				\\}

			\author[0000-0002-4971-5854]{Erico L. Rempel}
			\affiliation{Department of Mathematics, Aeronautics Institute of Technology, Pra\c ca Marechal Eduardo Gomes 50, 12228-900, S\~ao Jos\'{e} dos Campos, SP, Brazil \\
			}
			
			\author[0000-0002-9546-2368]{Gary Verth}
			\affiliation{Plasma Dynamics Group, School of Mathematical and Physical Sciences, University of Sheffield, Sheffield S3 7RH, UK
				\\}
			
			\author[0000-0002-3066-7653]{Istvan Ballai}
			\affiliation{Plasma Dynamics Group, School of Mathematical and Physical Sciences, University of Sheffield, Sheffield S3 7RH, UK
				\\}
			
			
			\author[0000-0002-0893-7346]{Viktor Fedun}
			\affiliation{Plasma Dynamics Group, School of Electrical and Electronic Engineering, University of Sheffield, Sheffield, S1 3JD, UK
				\\}

			
			
			
			\begin{abstract}
				Solar vortex regions show enhanced Poynting flux and favourable heating conditions, but how the vortices reorganize and influence their surroundings remains unclear
				Here we apply information-theoretic diagnostics to a Bifrost simulation to quantify the dynamics of a long-lived vortex.
				By combining Shannon Entropy and Normalized Mutual Information, we track how the vortex reshapes plasma-magnetic couplings and modifies local thermodynamics.
				The vortex originates in the upper photosphere and extends into the chromosphere, where it suppresses the background p-mode-like organisation seen in the neighbouring magnetic flux tube.
				Shannon Entropy analysis shows that magnetic complexity rises sharply as the vortex develops, which is consistent with the build-up of currents and stored energy. At the same time, temperature becomes more strongly linked to magnetic shear, pointing to heating associated with current dissipation.
				The way temperature responds to different heating processes also changes with height: in the photosphere and lower chromosphere, it follows local compressional and expansion motions, while in the upper atmosphere, it is influenced mainly by viscous and current-driven effects. During this phase, the usual temperature-density relationship weakens, indicating that the plasma departs from purely adiabatic behaviour.
				Applying the same diagnostics to a nearby non-vortical flux tube yields only weak, uniform couplings, which confirms that the enhanced links are vortex-driven.
				Together, these results demonstrate that a coherent solar vortex not only drives heating but also reconfigures the local atmosphere, replacing periodic pressure-driven behaviour with magnetically dominated dynamics.

			\end{abstract}

			
			\section{Introduction}
			
			The solar atmosphere supports a variety of vortices, from small-scale along intergranular lanes \citep{Giagkiozis_2018,Silva_2020,Yuan2023,Aljohani_2022,Shelyag_2011} to large vortical structures observed at the supergranular junctions \citep{Requerey_2017, Silva_2018, Chian_2019}. The intergranular vortices may hit the upper atmosphere, displaying larger rotational signatures in the CaII and H$\alpha$ core lines \citep{Wedemeyer2012,Dakanalis_2021,Dakanalis_2022} known as chromospheric swirls. 
			\cite{Wedemeyer2012} traced observational signatures of vortices across the solar atmosphere, showing that they form structures often called solar tornadoes. \cite{Silva_2024a} identified that solar tornadoes are, in fact, composed of three co-spatial vortices: a classical flow vortex \citep[the kinetic vortex,][]{Silva_2021}, a magnetic vortex~\citep{Silva_2021} consisting of coherent twisted magnetic fields, and a Poynting flux vortex, characterized by swirling trajectories of energy flow~\citep{Silva_2024a}.
			
			Numerical simulations have revealed a range of physical processes operating within vortex tubes. In high plasma-$\beta$ regions, plasma flows can twist magnetic fields \citep{Silva_2021}, generating magnetic shear and launching torsional waves \citep{Fedun2011, Yadav_2021}. In contrast, in low plasma-$\beta$ chromospheric regions, the dynamics can become magnetically dominated, with the magnetic field exerting control over plasma motions through magnetic tension and Lorentz forces \citep{Silva_2022,Silva_2024b}. Several studies have also highlighted the development of thin current sheets and localized heating at vortex boundaries, suggesting that magnetic shear and current dissipation may be central to vortex-related heating \citep{Yadav_2021, Silva_2021,Silva_2024a,Kuniyoshi_2025}. Vortices have also been linked to increased electromagnetic energy and plasma heating \citep{Kuniyoshi_2023, Kuniyoshi_2024, Silva_2024b, Silva_2024a, przybylski2025}. 
			
			Previous studies have shown that vortices in the solar atmosphere often coincide spatially and temporally with magnetic field concentrations, plasma heating, shocks, and jet formation \citep[e.g.,][]{Requerey_2017, Tziotziou_2023, Diaz_2024}. However, the nature and extent of these associations remain largely qualitative, leaving a gap in quantitatively characterizing their physical relationships. That is, they identify correlations based on co-location or apparent simultaneity, but they do not establish whether the vortex directly drives the observed dynamics. For instance, in the long-lived supergranular vortex studied by \citet{Requerey_2017, chian_2023}, the intensification and stabilization of a magnetic network element were observed to occur alongside vortex formation, but no formal causal analysis was performed. Consequently, these studies leave open a key question of whether vortices cause heating and magnetic structuring, or merely co-evolve with them.
			
			In this study, we use Information Theory (IT) methodology to investigate a solar tornado captured in a 3D simulation of a coronal hole. Our goal is to go beyond visual associations and directly measure how key physical variables interact over time. IT tools, though only recently introduced to space plasma studies, have proven transformative in uncovering hidden structure and causality in nonlinear systems. Unlike traditional correlation-based diagnostics, IT methods can identify directionality and time-lagged interactions in turbulent multiscale environments. For example, transfer entropy has been used to show that solar flares may influence the timing of subsequent flaring events \citep{Snelling_2020}, and to quantify how solar wind variability drives magnetospheric and radiation belt responses \citep{Wing2019}. More recently, IT has revealed vortex-flow coupling in turbulent plasmas \citep{Yatomi2025}, capturing causal pathways and scale-to-scale information transfer missed by conventional approaches. 
			
			By computing both Shannon Entropy (SE) and Normalized Mutual Information (NMI), we can quantify, for the first time, the degree of complexity of the plasma dynamics in the region where a solar tornado is present and also the strength of the coupling between vorticity and key MHD variables during the vortex flow lifecycle. To validate our results, we performed the same analysis focusing on a region with a magnetic flux tube without the presence of vortical flow. This multi-method approach allows us to address several open questions: (i) What mechanisms dominate the thermal evolution within large-scale solar vortices? (ii) How do magnetic shear and vorticity evolve during vortex formation and decay? (iii) Are heating processes localized and dynamic, or diffuse and passive? (iv) Critically, can we separate correlation from causation in vortex-driven energy transfer? 
			Through information-theoric lens, we provide in this paper a novel way to address such critical questions.
			The methods presented here provide a general framework for applying information theory to both simulations and observations of the solar atmosphere, offering novel tools for diagnosing energy transport in highly nonlinear and data-limited environments. The paper is organized as follows. In the Methodology section, we introduce the numerical simulation used in our analysis, together with the mathematical framework of the IT methodology and the magnetic flux-tube tracking procedure. Our analysis of the dynamical complexity and coupling in the vortical region is presented in the Results section. Finally, in the Discussion and Conclusions, we further interpret our findings and summarize the main implications of this work.

			\section{Methods}
			
			\subsection{Numerical Model: Bifrost simulation}
			
			The simulation employed in this study is the dataset \texttt{ch024031\_by200bz005}, produced using the \texttt{Bifrost} code \citep{Gudiksen_2011}.  The \texttt{Bifrost} code is a 3D  resistive, radiative magnetohydrodynamics (MHD) model that captures the coupled dynamics of the solar convection zone, photosphere, chromosphere, transition region, and lower corona within a single, self-consistent framework. It employs sixth-order finite differences for spatial derivatives on a staggered mesh and third-order time integration with an adaptive time-step. The Bifrost model incorporates several key physical effects, including thermal conduction along magnetic field lines \citep{Spitzer1957}, Ohmic heating, viscous dissipation, and a comprehensive set of radiative heating and cooling processes. To manage numerical diffusion, Bifrost employs a split-diffusive scheme that separates the diffusive terms into a small, globally applied component and a localized hyperdiffusive component.
			
			The analysed dataset consists of a vertically extended, low-density corona with a magnetically open topology. Thus, it captures key signatures of coronal holes, including reduced EUV emission and vertical magnetic connectivity. This simulation has also been used in earlier studies to examine vortex dynamics \citep{Silva_2024a, Finley_2022}. The simulation domain spans $24 \times 24 \times 16.8$~Mm$^3$, resolved with $768^3$ grid points, and extends from $2.5$~Mm below the solar surface (defined at optical depth unity, $z = 0$~Mm) up to $14.3$~Mm into the corona. The lower boundary includes a steady horizontal magnetic inflow, injecting a uniform field of 5~G, superimposed on a net vertical field of 200~G. The simulation setup creates a strongly unipolar magnetic environment with open field lines. The upper boundary is open, and the horizontal boundaries are periodic. The energy balance is handled through full radiative transfer in the lower atmosphere. This includes non-LTE scattering \citep{Carlsson_2012},  optically thin radiative losses, anisotropic thermal conduction, and ambipolar diffusion. 
			
\begin{figure*}[htbp]
				\centering
				\includegraphics[trim=0 100 0 0, clip, width=\textwidth]{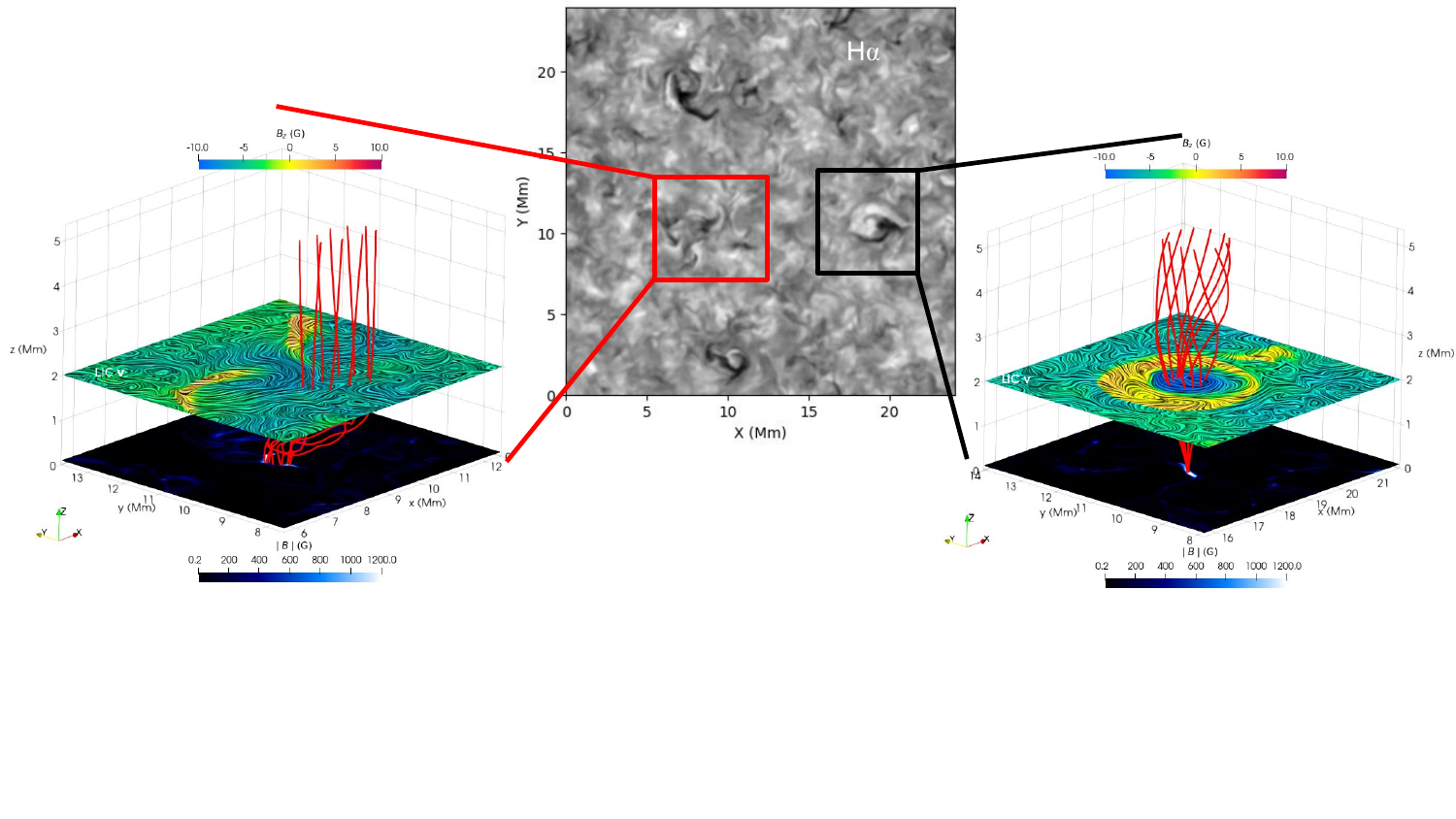}
				\caption{\textbf{Central panel:} Horizontal extent of the \textsc{Bifrost} simulation domain at \( t = 500 \) s, showing the  synthetic H$\alpha$ core line. The two analyzed subregions are marked by the red and black squares, respectively. 
					\textbf{Left panel:} 3D view of the region without large-scale vortical activity. Red lines indicate selected magnetic field lines seeded near the centre of the subdomain. The horizontal slice at \( z = 2 \)~Mm is colored by \( B_z \), with line-integrated convolution (LIC) patterns overplotted to reveal the horizontal velocity field. The bottom slice at the photosphere (\( z = 0 \)~Mm) also shows \( B_z \) for context. 
					\textbf{Right panel:} Same as left, but for the region containing the dynamically tracked large-scale kinetic vortex. The LIC texture highlights the coherent rotational motion, and the magnetic field lines reveal a twisted, vertically extended structure anchored in the vortex core.
				}
				\label{fig:vortices}
			\end{figure*}

			From the full simulation domain, we selected two sub-regions of \(6 \times 6\)~Mm\(^2\) on the horizontal plane. One region hosts a solar tornado. This structure has been previously analyzed in detail by \cite{Silva_2024a}. The second region contains only a vertical magnetic flux tube and serves as a control region, used as a reference to compare with the vortex case, as it exhibits no dominant vortical motion. To place the vortex analysis in spatial context, Figure~\ref{fig:vortices} shows the full horizontal extent of the \textsc{Bifrost} simulation colored by the synthetic H$\alpha$ core line. The spectra were computed accounting for departures from local thermodynamic equilibrium (non-LTE). To reduce the computational expense, we used \textsc{SunnyNet} \citep{Chappell2022}, which employs convolutional neural networks to approximate non-LTE hydrogen populations. The network was trained on two simulation snapshots generated with \textsc{Multi3D} \citep{Leenaarts_2009}.
			The two selected subdomains are indicated by the black and red squares. 
			The black square marks the region containing the large-scale, long-lived vortex, 
			while the red square denotes a nearby control region. 
			In the close-up views, the $xy$-planes are colored by the vertical component of the magnetic field. 
			The photospheric field tends to be highly concentrated at the footpoints of magnetic flux tubes, 
			leading to a dark colour almost everywhere else. The flux tubes are indicated by the magnetic field lines in red. These contrasting subdomains allow us to assess the influence of vortex dynamics on the evolution of magnetic structures and plasma variability.

			\subsection{Dynamic Masking and Spatial Focus}
			
			To isolate the dynamics local to the kinetic vortex, we applied a spatial mask that automatically tracks the evolving flux tube through height and time. The mask was defined as a circular region of fixed physical radius (in Mm) 
			applied to each horizontal $xy$-slice of the vortex region. This mask served to limit all statistical calculations, namely SE and NMI, to the immediate neighborhood of the vortex core. Although the vortex radius varies with height, the IT analysis required the regions to have the same size. This led to a compromise: an underestimation of the vortex region in the upper atmosphere and an overestimation near the photosphere. Nevertheless, the size of the analyzed region remains broadly consistent with the vortex extent for most of its vertical range.
			
			The mask centre was automatically determined for each height and time step as follows:
			
			\begin{enumerate}
				
				\item At each height, we identified the location of maximum magnetic field strength ($|\bf{B}|$) within the selected field of view. This point marks the central axis of the magnetic flux tube.
				
				\item We placed a circular mask of fixed radius centered on this point. To capture both the photospheric and chromospheric parts of the vortex, each with different horizontal sizes, we chose a compromise radius of 0.5 Mm that ensures consistent coverage across heights.
				
				\item To keep the mask vertically aligned, we limited the displacement of its centre to within 1~Mm of its position in the adjacent height layer. At our spatial resolution (0.03125~Mm per pixel), this corresponds to a maximum shift of about 32 pixels. Candidate centres were first searched within a larger neighborhood (12~Mm radius) to avoid losing the structure, but any displacement exceeding the 1~Mm limit was rejected, and the previous centre was retained. This procedure prevents the mask from locking onto nearby structures or noise. We used the same displacement constraint for all time steps. This is justified by the short cadence (10 seconds) and the small horizontal drift of the vortex
				
			\end{enumerate}
			
\begin{figure}
				\centering
				\includegraphics[width=\linewidth]{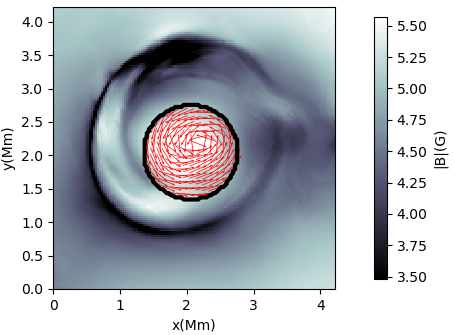}
				\caption{Magnetic field strength  $|\mathbf{B}|$ (grayscale, in Gauss) with unit velocity vectors (red) shown inside the circular mask (black contour) that automatically identifies the magnetic flux region at $z=3$ Mm and $t=500$ s.
				}
				\label{fig:masked}
			\end{figure}

			We ran this tracking method automatically across all time steps and heights, effectively building a 3D cylindrical region that follows the magnetic flux tube over time.  Magnetic flux concentrations appear before the vortex forms and remain after it fades away, making them a more consistent reference. This approach also lets us compare regions with and without strong vortex activity to better isolate the effects of coherent rotation. An example of the selected region at 500 seconds and 3 Mm height is shown in Figure \ref{fig:masked}, with red arrows marking the direction of the velocity field and the location of the mask.
			
			\subsection{Information Theory Quantification}

			To investigate the variability and statistical relationships among the plasma quantities within the vortex region, we applied IT tools to several variables, including the vertical magnetic field ($B_z$), total magnetic field strength $|\mathbf{B}|$, vertical velocity ($v_z$), vorticity, temperature ($T$), density ($\rho$), and pressure ($P$). In the simulation domain, these quantities exhibit strong height dependence; therefore, within the masked vortex region, each variable was normalized to enable meaningful comparison across time and height. Because IT measures need discrete probability distributions, we converted each continuous field into a histogram within the mask.

			To compute SE, the range of each variable was divided into \( K \) bins, and the empirical probability, \( p_i \), was determined as the fraction of grid cells with values falling into bin \( i \), such that
			\[
			\sum_{i=1}^{K} p_i = 1.
			\]
			The entropy was then calculated in bits (SE units) using, 
			\begin{equation}
				H = -\sum_{i=1}^{K} p_i \log_2 p_i .
			\end{equation}
			To determine the number of bins, $K$, we tested both adaptive and fixed binning approaches. Adaptive binning adjusts to the local spread of the data, but it often over-resolves the histogram in regions with narrow distributions or limited sample sizes. This resulted in nearly uniform probability distributions and artificially high entropy values, which washed out spatial and temporal contrasts in the entropy maps. Fixed binning, on the contrary, preserved coherent structures and transitions across all variables and more clearly reflected known dynamic processes, such as vortex emergence and decay. For this reason, we adopt the fixed binning as the more interpretable and physically consistent method.
			
\begin{figure*}
				\centering
				\includegraphics[width=1\linewidth]{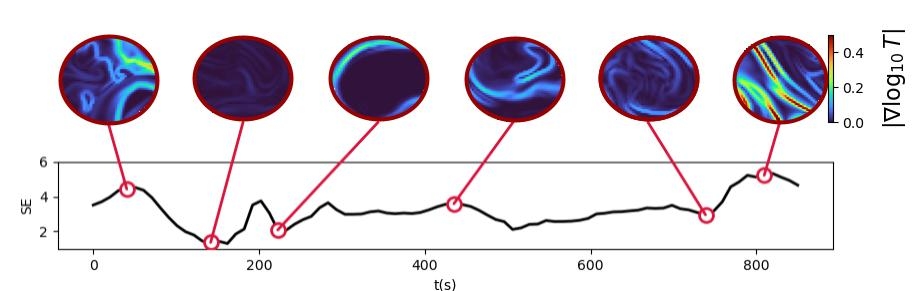}
				\caption{Temporal evolution of the SE computed from the $\log_{10}$ of the plasma temperature within the masked region of the magnetic flux tube, at a height of 2.52~Mm, where a vortex develops. The upper circular panels display snapshots of the temperature gradient magnitude, $|\nabla \log_{10} T|$, at selected times along the entropy curve. 
				}
				\label{fig:SE_interpret}
			\end{figure*}

			Mathematically, the SE measures the level of uncertainty associated with a probability distribution. If we compute the SE of a narrow Gaussian distribution where most values fall within a small range, the SE value will be low, as the outcome of a random sample can be predicted with high confidence. On the other hand, if the distribution is broader or flatter, as in a wide Gaussian, the probability is more evenly spread across a larger range of possible values, making it harder to predict the outcome, resulting in a higher entropy. In our analysis, SE of a 2D array quantifies the spatial variability or disorder of a physical quantity. A high entropy value indicates strong spatial fluctuations, related to a complex or possibly turbulent structure. In contrast, low entropy implies greater spatial uniformity, reflecting a more coherent or stable region. The lower panel of Figure~\ref{fig:SE_interpret} shows the SE computed at a height of 2.52~Mm for the plasma temperature within the magnetic flux tube that exhibits rotational plasma flow. For selected times, the upper panel of 
			The lower panel of Figure~\ref{fig:SE_interpret} shows the SE computed at a height of 2.52~Mm for the plasma temperature within the magnetic flux tube that exhibits rotational plasma flow. For selected times, the upper panel of Figure~\ref{fig:SE_interpret} displays the magnitude of the temperature gradient, \(|\nabla \log_{10} T|\). Figure~\ref{fig:SE_interpret} illustrates that higher SE values (above 4) generally correspond to stronger and more irregular temperature distributions, as reflected in the corresponding temperature gradient maps. In contrast, lower SE values (equal to or below 2) are associated with more homogeneous temperature fields and correspondingly weaker gradients.
			The same reasoning applies to the other physical variables analyzed, where higher SE values indicate greater spatial complexity, higher gradients, and dynamical activity. For instance, the physical interpretation of high entropy in $B_z$ or $|\mathbf{B}|$ suggests that the magnetic field is highly structured, possibly due to twisting, braiding, or the formation of current sheets. The low entropy in these quantities indicates an organized field, as seen in magnetically dominated regions or within stable flux tubes. For velocity and vorticity, high entropy reflects disordered or fine-structured flows. This can arise from turbulent motions or from overlapping wave modes, such as Alfvén or magnetoacoustic waves, which introduce spatial variability as they propagate through the magnetized plasma. When applied to thermodynamic variables such as density and pressure, high entropy indicates significant spatial gradients and inhomogeneity, often associated with localized shocks or dynamic flows. Conversely, low entropy in these fields suggests more uniform conditions, typical of equilibrium or quasi-static regions. The SE, thus, provides a quantitative diagnostic of the degree of structural organization and mixing within the magnetic environment, offering a more informative and objective measure for assessing how vortex dynamics influence the thermal and magnetic structuring of the plasma.
			
			\subsubsection{Quantifying statistical dependence with mutual information}
			
			MI is built upon the concept of entropy and provides a quantitative measure of the statistical dependence between variables.
			In both physics and information theory, entropy represents uncertainty or disorder, and MI quantifies how much knowledge of one variable reduces the uncertainty of another. In this sense, MI measures coupling strength, the degree to which two quantities share information. Unlike correlation, MI does not assume a specific functional form of the relationship, making it sensitive to both linear and nonlinear dependencies. 
			Since MHD system are inherently nonlinear, traditional correlation-based measures may miss subtle, yet important dependencies. MI, however, is sensitive to the full spectrum of possible interactions as it stems from entropy, a concept central to both thermodynamics and information theory.
			
			To compare how strongly different pairs of variables are connected, or how this connection changes across heights or times, MI is often normalized. This places all variable comparisons on a common scale, making it possible to assess the relative strength of statistical dependencies in a consistent manner. Several normalization schemes exist, such as dividing MI by the joint entropy or by the geometric mean of the individual entropies. In this study, we adopt the latter form,
			\[
			\mathrm{NMI}(X,Y) = \frac{I(X;Y)}{\sqrt{H(X)H(Y)}},
			\]
			which ensures symmetry between variables and yields values between 0 and 1. The geometric-mean normalization is a statistical choice introduced to enable comparison across variable pairs; it does not carry a direct physical meaning and may smooth out possible asymmetries in directional couplings. Once normalized, however, the resulting NMI values can still be interpreted in physical terms, as they express the fraction of shared information between two variables. For instance, an NMI value of 0.1 indicates that the two MHD variables share about 10\% of their total information content.
			
			In other words, knowing the state of one variable reduces the unpredictability of the other by that fraction.
			In practical terms, a normalized MI above 0.1 indicates a nontrivial statistical dependency. For MHD, this can reveal meaningful coupling between physical processes, such as energy transfer across fields, cross-scale interactions in turbulence, or correlations between magnetic and velocity structures that would not necessarily be visible in linear measures. Thus, even a seemingly “small” MI value may reflect deep physical connections within the plasma system, encompassing both dynamical and thermodynamic relationships.			
			For two random variables $X$ and $Y$, their MI, $I(X;Y)$, is defined as
			\begin{equation}
				I(X;Y) = H(X) + H(Y) - H(X,Y).
			\end{equation}
			Here, $H(X, Y)$ denotes the joint entropy between variables $X$ and $Y$, which quantifies the total uncertainty associated with their combined outcomes. It reflects not only the individual variability of each variable, but also its shared structure. We normalized $I(X;Y)$ to account for differences in the marginal entropies of $X$ and $Y$. The normalised MI (NMI) is defined as
			\begin{equation}
				\mathrm{NMI}(X,Y) = \frac{I(X;Y)}{\sqrt{H(X) H(Y)}}.
			\end{equation}
			Before computing the NMI, each variable was rescaled to a fixed range, either \([0,1]\) or \([-1,1]\), depending on its physical bounds. Joint histograms were then constructed using 64 uniformly spaced bins across the normalized range. This fixed binning approach provides enough resolution to capture nonlinear dependencies while maintaining numerical stability in the mutual information estimates. Normalizing and binning in this way ensures consistency across variables and time steps, allowing for reliable comparisons and minimizing sensitivity to outliers or local fluctuations in the data.
			To assess the statistical reliability of the observed information-theoretic coupling, quantified by the NMI, we performed pixel-shuffle surrogate tests within each time-height window. For each driver-target variable pair (e.g., temperature as the driver and velocity or magnetic field components as the target), we generated \( N = 500 \) spatial surrogates by randomly permuting pixel positions within the vortex mask. This procedure preserves the marginal amplitude distributions of each variable while destroying their spatial correspondence, thereby eliminating genuine statistical dependencies between them. Using the surrogate ensemble, we computed the MI values to construct a null distribution, from which the significance level, \(p_{\mathrm{sig}}\), was defined as the fraction of surrogate samples with MI values greater than or equal to the observed one. Across the entire analysed region and time interval, \(p_{\mathrm{sig}} \lesssim 0.002\) in nearly all pixels, with only isolated exceptions.
			
			\section*{Results}
			
\begin{figure*}
				\centering
				\includegraphics[width=1\linewidth]{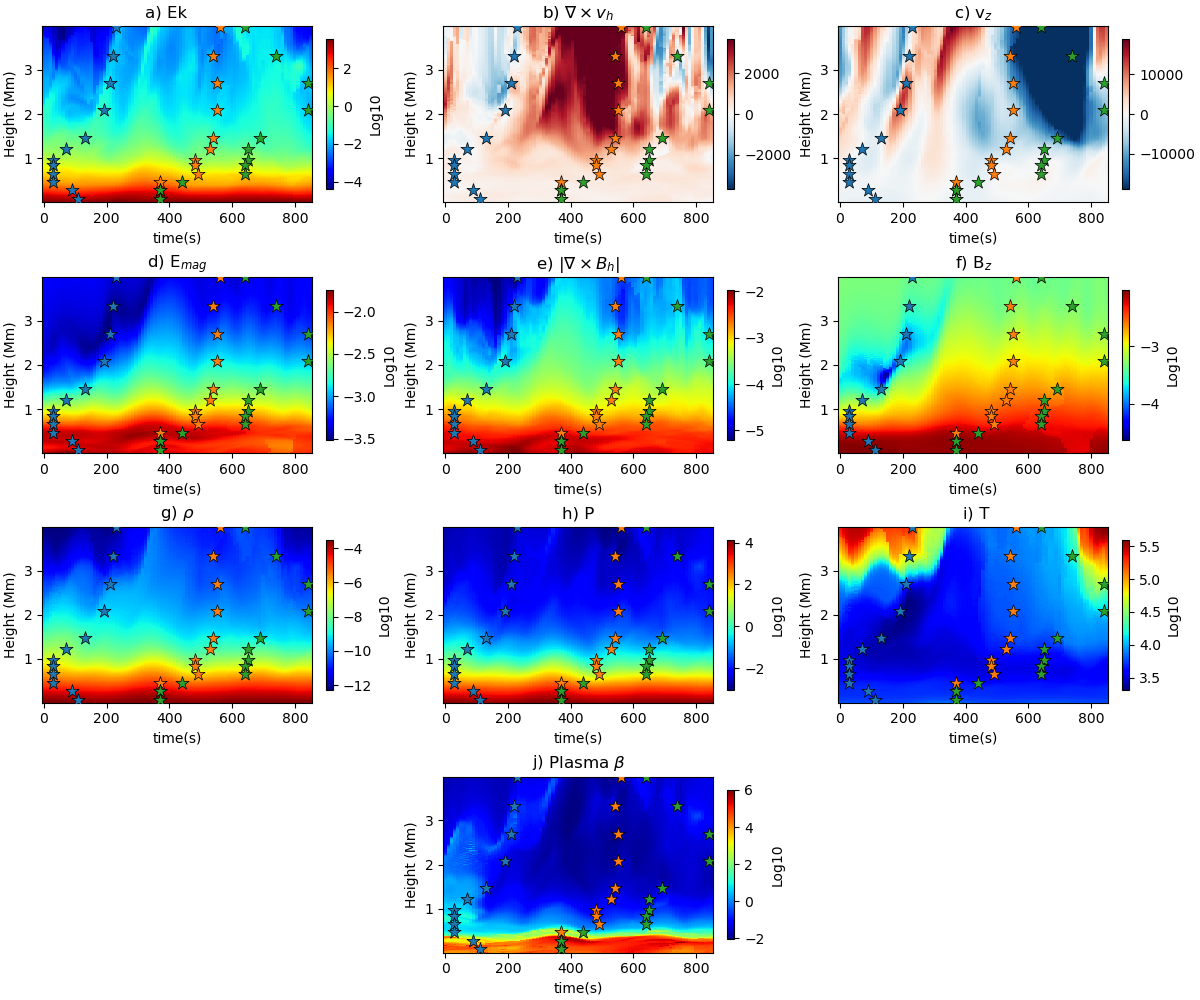}
				\caption{Temporal and vertical evolution of the spatially averaged plasma quantities within the dynamically tracked magnetic flux core. Each panel corresponds to a different physical variable: a) kinetic energy  $Ek$, b) vorticity of horizontal velocity $\nabla \times \mathbf{v}_h$, c) vertical velocity $v_z$, (d) magnetic energy density $E_{\text{mag}}$, e)  magnitude of magnetic shear $\left| \nabla \times \mathbf{B}_h \right|$ of the transverse magnetic field, f) vertical magnetic field $B_z$, g) mass density $\rho$, h) gas pressure $P$, i) temperature $T$, and j) plasma-$\beta$. All variables are shown in base-10 logarithmic scale, except for the vorticity $\nabla \times \mathbf{v}_h$ of horizontal plasma flows and the vertical velocity $v_z$.
					Blue stars mark the onset of the main large-scale vortex, solar tornado, at each height, while orange stars indicate its decay. Green stars denote the termination of smaller secondary vortices that form after the decay of the primary structure. All times and locations were identified using the $\Gamma$-method with $\Gamma>1$. 
				}
				\label{fig:variables}
			\end{figure*}

			To characterize the temporal and spatial (in vertical direction) evolution of physical conditions within the vortex, we computed time-height maps of spatially averaged quantities inside the tracked magnetic flux tube core. By analyzing these averaged fields across the different vortex phases, before onset, development, and decay, we identify the signatures of dynamic coupling and energy concentration associated with vortex-driven activity. Figure~\ref{fig:variables} shows the temporal and height variation of the spatially averaged vorticity of horizontal velocity ($v_h$), vertical velocity ($v_z$) and the base-10 logarithm of each of the other variables within the magnetic flux tube. The spatial averages were computed for each height plane. The stars of different colours indicate key stages in the vortex evolution as determined by the $\Gamma$-method \citep{Graftieaux2001,Giagkiozis_2018, Yuan2023}, using the condition $\Gamma_1 > 0.66$ in the magnetic flux tube mask to define coherent rotation. We choose a lower value as initially the velocity field is not very twisted, leading to a lower $\Gamma_1$ value. Blue stars mark the onset of the main vortex at each height, while orange stars indicate the end of it. As the main vortex decays, it fragments into smaller secondary vortices that persist for a time before fully dissipating; their presence is marked by green stars. The vortex appears to form between the high photosphere (approximately 0.45 Mm) and the low chromosphere (around 1.0 Mm), after which it expands both upward into the corona and downward toward the solar surface. In the upper atmosphere, the vortex decays into multiple smaller structures, whereas in the lower layers it disappears more abruptly. In particular, smaller vortices are more persistent around 2 Mm, indicating that vortex evolution may be stratified with height, with different lifetimes and dynamics depending on atmospheric depth.

			The formation of the strong, coherent vortex leaves clear signatures in all plasma variables, especially in the chromosphere and transition region (above roughly 1 Mm). In the upper atmosphere, the vorticity changes sign around the time the vortex forms. This marks a shift from small, network-like flows that rotate clockwise to a larger, dominant vortex rotating counterclockwise. The positive vorticity linked to this large-scale vortex lasts through most of its lifetime but slowly weakens as the coherent rotation breaks apart. During the decay phase, the vortex transitions into smaller, fragmented swirling motions. This pattern shows that as the vortex develops, angular momentum spreads outward and energy shifts from the organized, large-scale rotation into smaller, irregular motions that can dissipate energy through heating.
			
			One of the most prominent features is a sharp and sustained drop in plasma-$\beta$, which falls below unity shortly after vortex formation and remains low (log$_{10} \beta \lesssim -1$) as long as coherent vortical structures persist. This behaviour marks a transition from a gas-pressure-dominated to a magnetically dominated regime. Simultaneously, both the kinetic and magnetic energy increase by over an order of magnitude, peaking near vortex’s half-life. A local minimum in the vertical magnetic field \( B_z \) coincides with the onset of the main vortex (blue stars). This may reflect either a redistribution of magnetic flux by the swirling plasma motions or the conversion of vertical field into horizontal components as the vortex twists and shears the local field. The subsequent increase in \( B_z \) indicates magnetic flux accumulation within the vortex core as it matures.
			
			The formation of the vortex is accompanied by a coherent upflow of plasma, which transports denser fluid from lower atmospheric layers into the vortex core. This upflow results in a localized enhancement of mass density and a simultaneous drop in temperature, consistent with adiabatic expansion and the lifting of a cooler, denser plasma. Throughout the lifetime of the vortex, upward velocities dominate within the core region, continuously transporting plasma from lower layers and sustaining the elevated density. This persistent upflow plays a crucial role in maintaining the mass loading of the vortex structure. Only toward the end of the vortex’s life does this trend reverse, with downflows gradually becoming more prominent. This culminates in a strong downward jet along the vortex axis during its final decay phase, particularly in the upper atmospheric layers. The gas pressure initially increases at heights above 1.5 Mm as the upflow develops, but subsequently drops as the vortex matures. Following the initial cooling phase, the plasma within the vortex core is gradually heated, reaching temperatures up to an order of magnitude higher than the onset vortex state. 
			
\begin{figure*}
				\centering
				\includegraphics[width=1\linewidth]{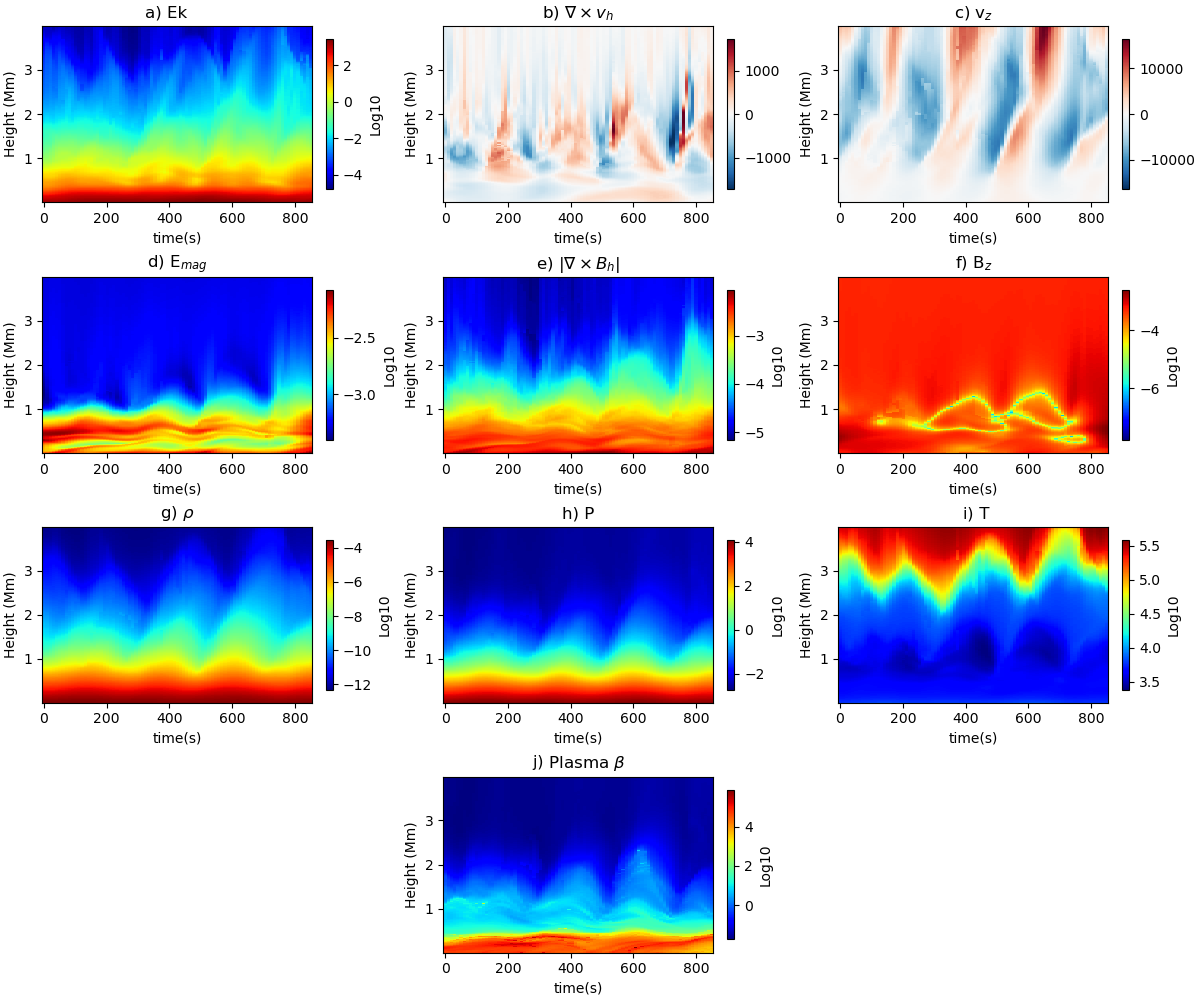}
				\caption{Same as Figure \ref{fig:variables}, but for the magnetic flux tube without coherent rotational motion.
				}
				\label{fig:novortex_variables}
			\end{figure*}

			Figure~\ref{fig:novortex_variables} displays the evolution of the entropy in a control region without coherent vortex activity. All variables exhibit nearly periodic fluctuations resulting from global p-mode oscillations driven by the lower boundary of the Bifrost simulation, which imposes perturbations to emulate solar acoustic modes.
			The non-vortex region exhibits a more periodic and stratified behaviour, dominated by pressure-driven dynamics rather than magnetic stresses, indicating a relatively quiescent and stable plasma environment. It presents weaker magnetic fields and higher plasma $\beta$ values reflecting limited magnetic coupling and reduced energetic activity.
			A comparison with Figure~\ref{fig:variables} highlights fundamental differences in physical and dynamical properties. The vortex region lacks the clear periodic patterns seen in the non-vortex case, instead showing more irregular and complex temporal behaviour. It is characterized by stronger magnetic fields and enhanced magnetic twisting, particularly in the upper atmospheric layers, indicating that its dynamics are magnetically dominated. Correspondingly, the plasma $\beta$ is significantly lower in the vortex region, confirming that magnetic forces play a leading role in its evolution. The vortex region also exhibits a denser and progressively hotter atmosphere toward later times, consistent with magnetic energy concentration and conversion into thermal energy. In contrast, the non-vortex region maintains a more periodic, stratified, and plasma-pressure-dominated behaviour, typical of a less magnetically active environment.
			
			\subsection{Tracking Plasma Coupling and Coherence via Information Theory}
			
\begin{figure*}[htbp]
				\centering
				\includegraphics[width=\textwidth]{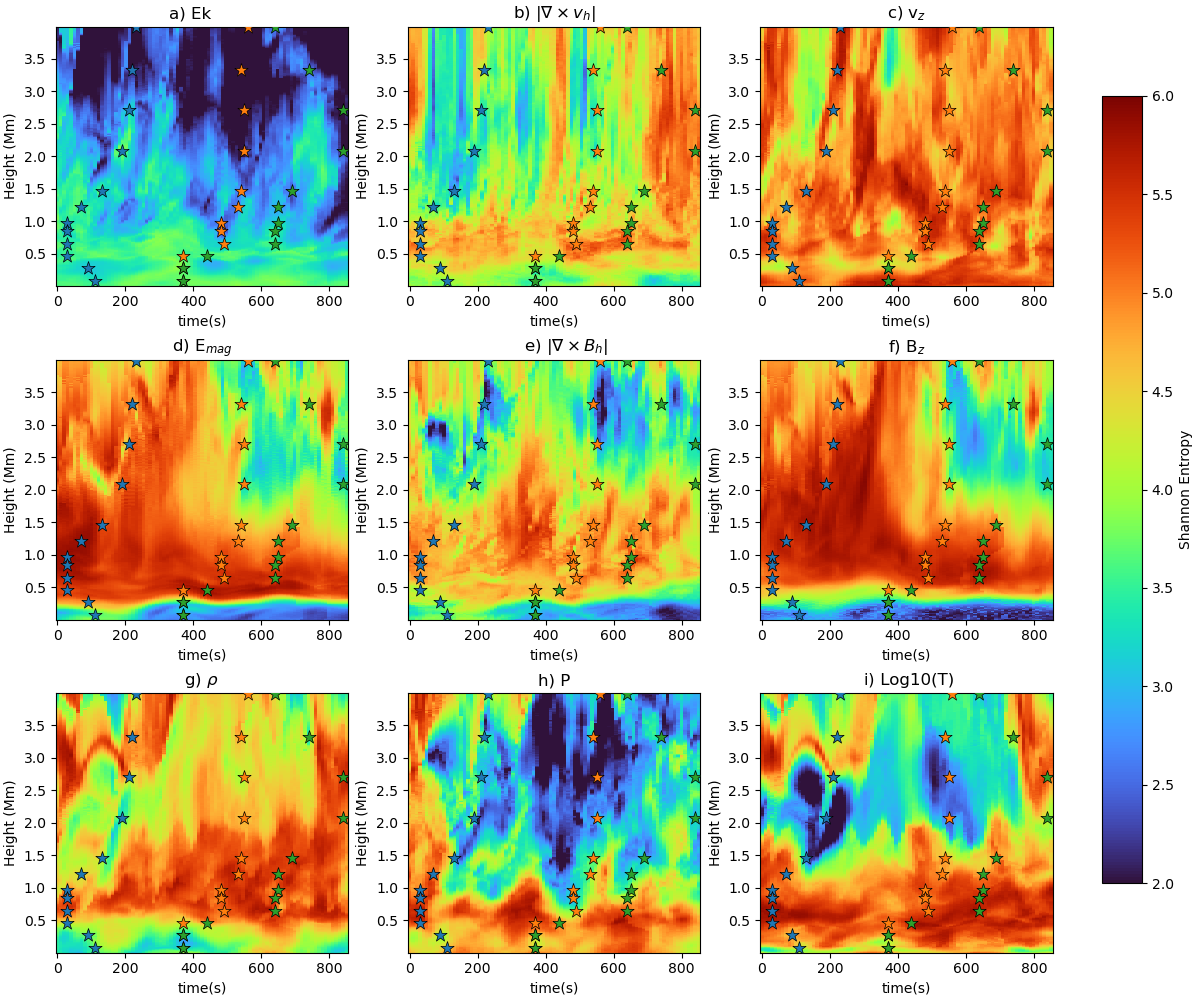}
				\caption{Temporal and vertical evolution of SE for various plasma variables, computed within the dynamically tracked vortex core. SE is estimated from the empirical probability distribution of each variable at every height and time, using fixed binning (64 bins). Each panel corresponds to a distinct quantity: a) kinetic energy density, $Ek$, b)  vorticity $\nabla \times \mathbf{v}_h$ of horizontal velocity flows, c) vertical velocity $v_z$, d) magnetic energy density $E_{\text{mag}}$, e)  magnitude of magnetic shear $\left|\nabla \times \mathbf{B}_h \right|$ of the transverse magnetic field, f) vertical magnetic field $B_z$, g) mass density $\rho$, h) gas pressure $P$, and i) temperature $T$. Blue and orange stars mark the onset and end of the large-scale vortex, while green stars indicate the disappearance of secondary vortex structures. 
				}
				
				\label{fig:entropy_vortex}
			\end{figure*}

			The profiles presented in Figure \ref{fig:variables} characterize the background evolution of the vortex core and provide a reference context to interpret subsequent information-theory and structural analyzes. The computed SE quantifies the local spatial variability of key physical quantities within the tracked magnetic flux tube where the solar tornado is developed. Figure~\ref{fig:entropy_vortex} shows the temporal and vertical evolution of the SE for the same variables as in Figure \ref{fig:variables}, except for plasma-$\beta$. The SE was computed within the dynamically tracked magnetic flux tube, and the entropy was estimated from the empirical probability distributions at each height and time step using a fixed binning scheme of 64 bins and logarithms in base~2, corresponding to a maximum possible entropy of \( \log_2(64) = 6 \)~bits. Each panel in Figure~\ref{fig:entropy_vortex} presents the entropy values, revealing how the spatial organization and complexity of the plasma change over time and height. In practice, entropy values near $H \sim 2$--$3$ correspond to relatively ordered fields with low spatial variability, while intermediate values ($H \sim 4$--$5$) indicate partially structured states where coherent features coexist with moderate fluctuations. Higher values approaching $H \sim 6$ reflect strongly intermittent fields, likely associated with turbulence, sharp spatial gradients, or enhanced small-scale variability. 
			
			The onset and development of the main vortex are marked by significant changes in the spatial organization of magnetic fields, as captured by entropy diagnostics. During the formation phase, the entropy of magnetic energy ($E_{\text{mag}}$) rises in the lower atmosphere, indicating increasing spatial complexity and localized structuring of the field. This is accompanied by a divergence between the entropy of $E_{\text{mag}}$ and that of magnetic shear ($|{\nabla} \times \mathbf{B}_h|$), consistent with the formation of current sheets rather than uniform shear enhancement. While $E_{\text{mag}}$ becomes more intermittent due to twisting and compression by the vortex, magnetic shear entropy remains vertically patchy and fragmented, suggesting that complexity arises from thin, spatially confined features rather than widespread braiding. Similarly, the entropy of the vertical field component ($B_z$) increases during the vortex onset-likely reflecting enhanced intermittency as flux elements are swept into the vortex-but later drops sharply between 1.5-3~Mm once the coherent structure is established. This entropy reduction suggests that the vortex acts to stabilize and organize $B_z$, possibly through magnetic tension and vertical realignment. 
			Before the onset of the vortex, both the magnitude and entropy of vorticity of horizontal plasma flows remain low in the upper atmosphere. This indicates a quiescent, non-rotational flow regime with minimal spatial variation, consistent with an irrotational, magnetically dominated background. The absence of pre-existing shear or turbulence suggests that the vortex did not arise from the gradual accumulation of small-scale swirling motions or steady twisting of field lines. Instead, it likely formed abruptly in response to a localized, possibly impulsive, event, potentially triggered by plasma or magnetic instabilities. The entropy of the kinetic energy ($Ek$) increases with vortex onset, decreasing immediately after, reflecting the emergence of coherent flow patterns. However, this trend reverses near the vortex’s half-life, where entropy increases significantly. This delayed rise is likely driven by the emergence of shear flows that act to destabilise the coherent vortex and inject kinetic complexity into the system. Notably, this entropy growth in $Ek$ is followed by a corresponding increase in the entropy of vorticity, which had been suppressed during the organized vortex phase. The subsequent increase in vorticity entropy marks the transition to a fragmented regime dominated by small-scale, disordered vortical structures, consistent with the formation of secondary swirls.
			
			Temperature and pressure both exhibit an initially high entropy in the lower atmosphere, reflecting significant spatial variability before the vortex formation. As the vortex emerges, entropy in both quantities decreases. Notably, pressure entropy drops more rapidly than temperature entropy, indicating that the pressure field becomes ordered earlier, possibly due to the propagation of magnetoacoustic waves that redistribute pressure more efficiently than thermal energy. Following this initial organisation phase, the two variables diverge in behaviour. The pressure entropy remains low throughout the middle and upper atmosphere during the lifetime of the vortex, with only a modest increase near the footpoints and below 1~Mm. In contrast, temperature entropy begins to rise in the upper atmosphere as the vortex develops, pointing to the formation of fine-scale thermal structure likely driven by localised heating and dynamic mixing processes. This growing thermal complexity aligns with the evolution of plasma temperature shown in Figure~\ref{fig:variables}, where the vortex core heats significantly during its mature phase, while pressure remains relatively stratified and stable in the same region.

\begin{figure*}[htbp]
				\centering
				\includegraphics[width=\textwidth]{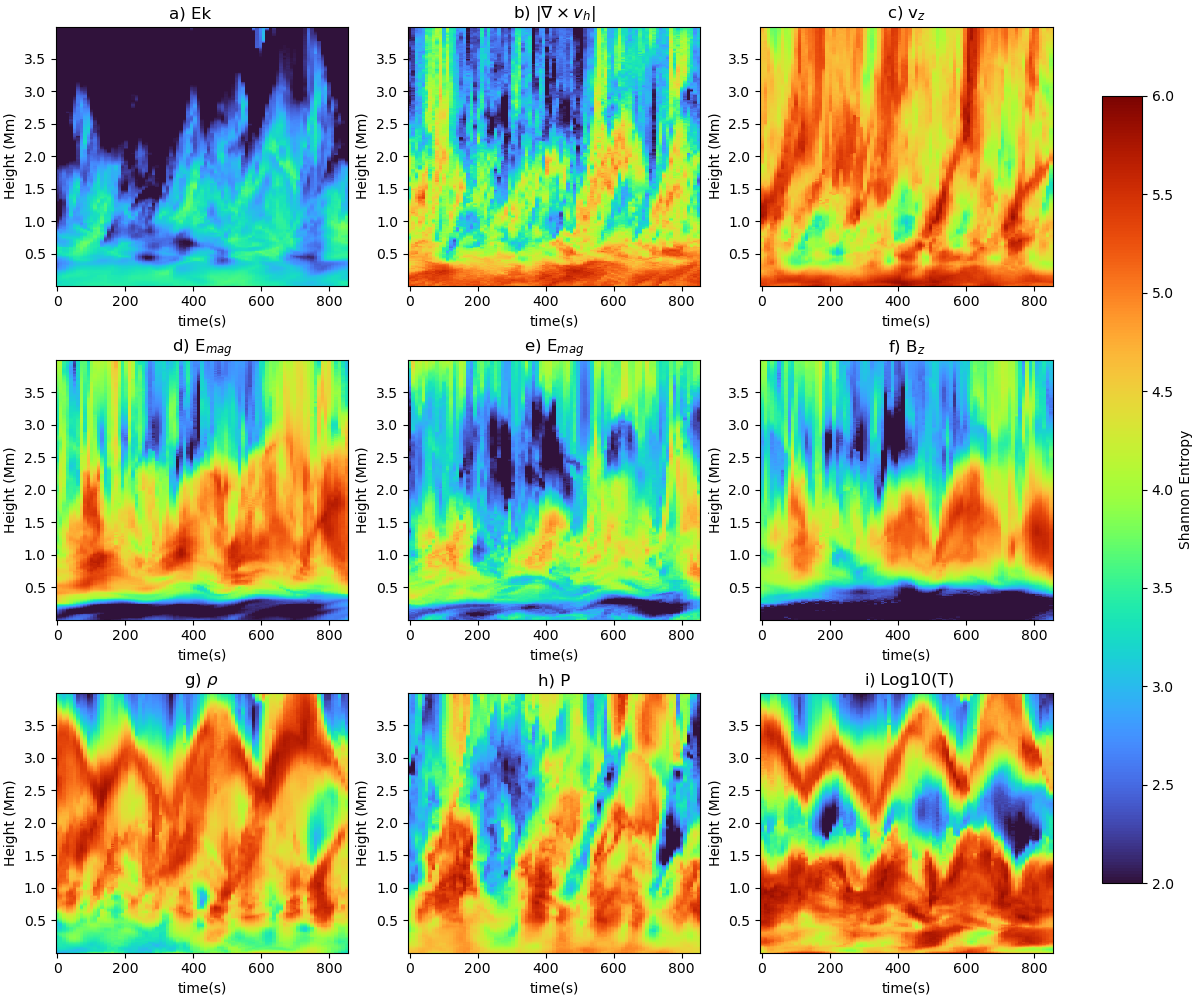}
				\caption{
					Same as Figure~\ref{fig:entropy_vortex}, but for a region without vortex activity. 
				}
				
				\label{fig:Entropy_novortex}
			\end{figure*}

			Figure~\ref{fig:Entropy_novortex} displays the evolution of the entropy in a control region without coherent vortex activity. A striking feature is the presence of vertically coherent, nearly periodic fluctuations across several plasma variables, particularly in density, pressure, and temperature. These patterns likely reflect global p-mode oscillations driven by the lower boundary of the Bifrost simulation, which imposes perturbations to emulate solar acoustic modes. These acoustic modes periodically compress and expand the plasma, modulating temperature, density, and pressure gradients. As a result, the spatial complexity of these quantities, measured through SE, oscillates in phase with the p-mode cycles. Thus, higher SE values correspond to the compressive phases with steeper gradients, and lower values to the rarefactive, smoother phases.
			Interestingly, such periodic signatures are suppressed in the vortex region (Figure~\ref{fig:entropy_vortex}), suggesting that the presence of the vortex alters or even decouples the propagation of global p-mode oscillations. The strong and twisted magnetic fields within the vortex modify the local wave environment, favouring the generation and propagation of magnetoacoustic modes confined within the magnetic structure. Previous analyses of the same simulation have identified such waves in the lower atmosphere, supporting the idea that the oscillatory behaviour within the vortex is primarily driven or modulated internally, by the vortex dynamics themselves, rather than directly inherited from the p-mode forcing at the lower boundary.

			\subsubsection{Flow-Rotation Coupling with Plasma Parameters}
\begin{figure*}[h!]
				\centering
				\includegraphics[width=\textwidth]{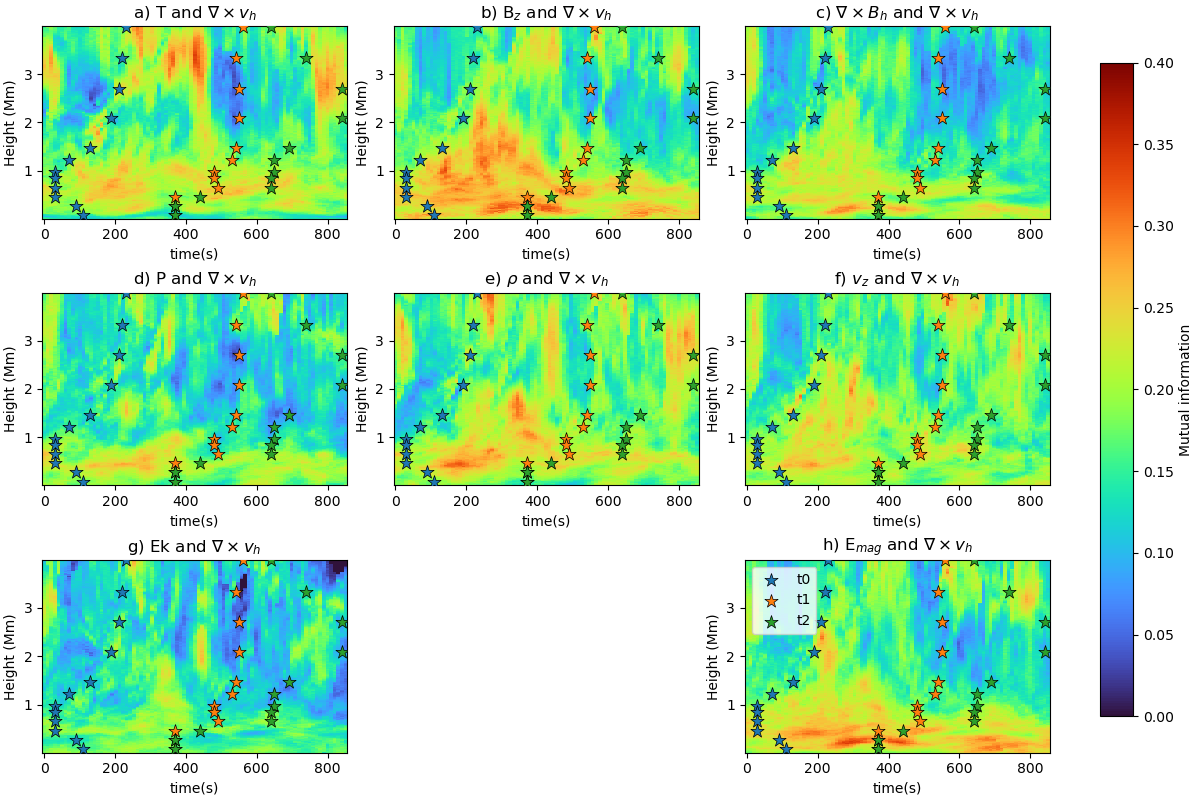}
				\caption{Temporal and vertical structure of the mutual information (MI) computed within the dynamically 
					tracked vortex core region. Different panels show MI between 
					$\nabla \times \mathbf{v}_h$ and:
					(a) temperature $T$,
					(b) vertical magnetic field $B_z$,
					(c) the magnitude of magnetic shear $ \left|\nabla \times \mathbf{B}_h \right|$ of transverse magnetic field,
					(d) gas pressure $P$,
					(e) density $\rho$,
					(f) vertical velocity $v_z$,
					(g) kinetic energy density $Ek$, and
					(h) magnetic energy density $E_{\rm mag}$.
					MI values are shown in normalized units (bits), with warmer colors
					indicating stronger dependence. Stars denote key moments in the vortex
					evolution: blue for onset ($t_0$), orange for decay of the main vortex
					($t_1$), and green for the dissipation of residual swirls ($t_2$).
					\label{fig:MI_vortex}}
			\end{figure*}

\begin{figure*}[h!]
				\centering
				\includegraphics[width=\textwidth]{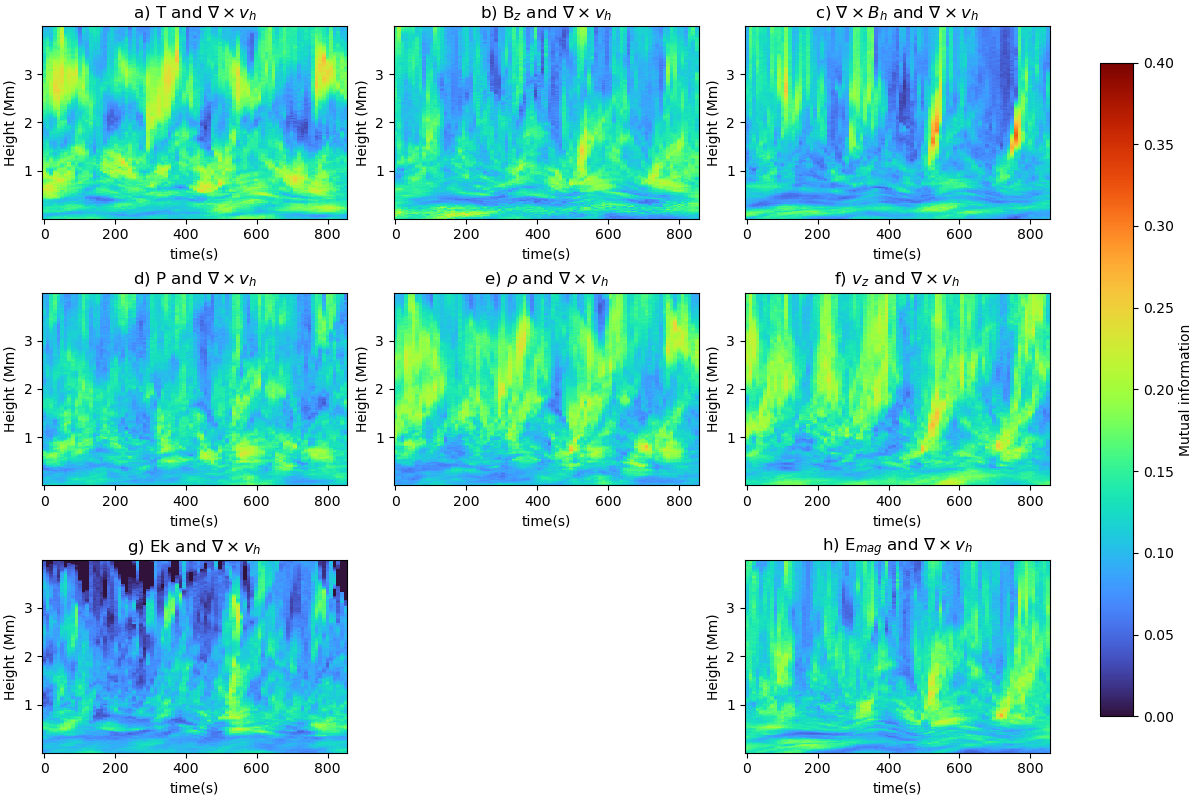}
				\caption{
					Same as Figure~\ref{fig:MI_vortex}, but for the control region where no coherent vortex is present, only a slightly twisted magnetic flux tube. 
					\label{fig:MI_novortex}}
			\end{figure*}

			To explore the degree of statistical dependence between rotational motion and key plasma parameters during the vortex lifecycle, we computed the NMI between the vorticity magnitude of horizontal plasma flow ($|\nabla \times \mathbf{v}_h|$) and each relevant physical quantity as displayed in Figure~\ref{fig:MI_vortex}. Here, MI values are normalized, so that MI$=0$ reflects no statistical dependence, while MI$=1$ denotes perfect dependence. NMI in Figure~\ref{fig:MI_vortex} ranges from 0 to approximately 0.4, providing a dimensionless measure of statistical dependence between vertical vorticity and each plasma variable. Values close to 0 indicate little to no shared information, reflecting largely uncorrelated or independent behavior. Moderate NMI values (~0.1-0.3) suggest partial coupling, where the variable exhibits some degree of organization influenced by the vorticity field, possibly through indirect or non-linear interactions. Higher NMI values ($\approx$0.3-0.4) indicate a stronger statistical dependence between variables, consistent with coordinated evolution.
			During the vortex phase, the coupling between vorticity and several thermodynamic and magnetic quantities increases markedly, particularly in the upper atmosphere.
			The control region without a vortex exhibits only weak and spatially uniform NMI patterns (Figure \ref{fig:MI_novortex}).
			This contrast shows that the enhanced coupling within the vortex core is a dynamic signature of vortex activity, not a static property of the background magnetic field.
			
			In the non-large-scale vortex region, where only small-scale swirling flows are present, NMI values remain generally low and spatially diffuse (Figure~\ref{fig:MI_novortex}) in the upper atmosphere. This suggests a weak statistical coupling between vorticity and other plasma variables when coherent, vertically extended rotational structures are absent. The elevated NMI between temperature, $v_z$, and vorticity in the photospheric control region likely arises from granular-scale convection, intergranular downflows, and p-mode-induced oscillations, all of which structure the plasma in ways that naturally correlate temperature, vertical motion, and rotational shear—even in the absence of a coherent vortex. Importantly, a low vorticity amplitude does not necessarily imply a low NMI, nor does a high vorticity guarantee strong coupling. What NMI captures is not the intensity of a given quantity, but the degree to which two variables co-vary in a structured and predictable manner. The persistently lower NMI in this control region reinforces the idea that strong mutual dependence, as seen in the vortex core, requires not just vorticity, but organized dynamics capable of modulating the surrounding plasma fields.

			\subsection{Plasma heating}
\begin{figure*}
				\centering
				\includegraphics[width=\linewidth]{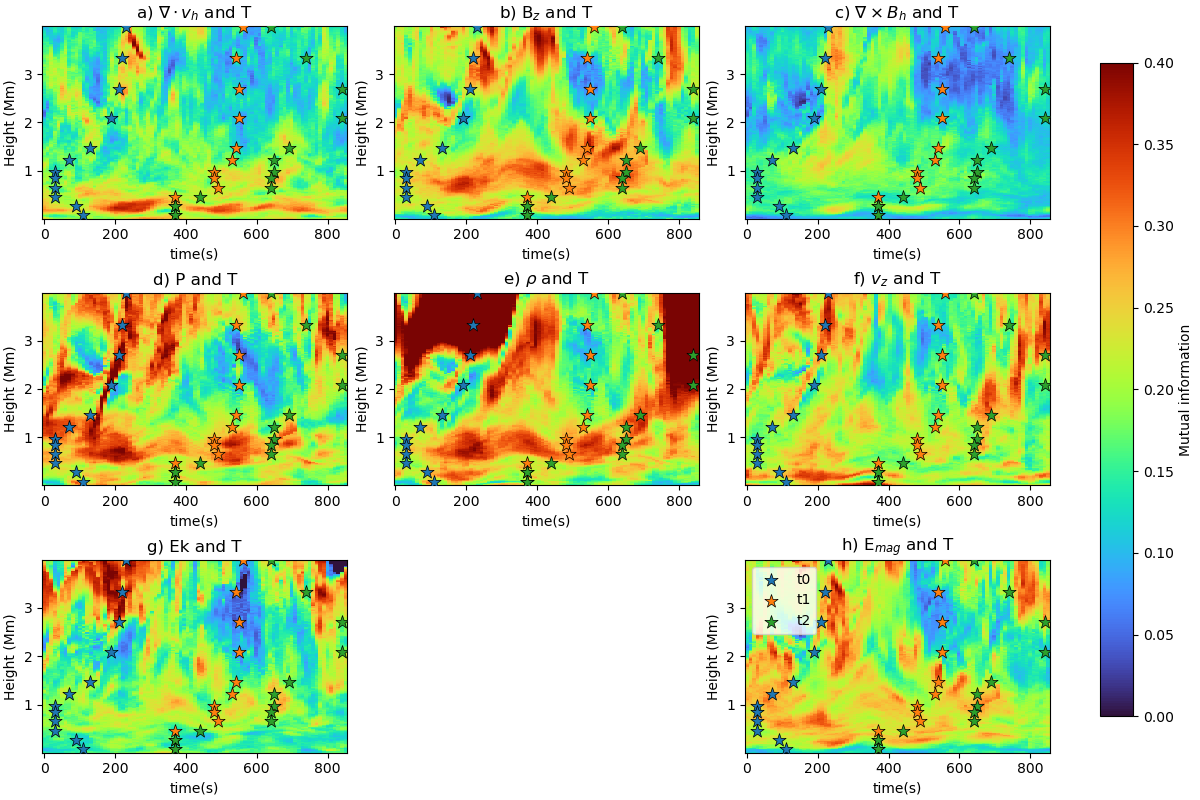}
				\caption{
					NMI between plasma temperature and other scalar quantities within the 
					dynamically tracked vortex core region. Each panel shows the NMI between 
					temperature and a different scalar variable: 
					(a) the velocity divergence of horizontal velocity field
					$ \nabla \cdot \mathbf{v}_h $, 
					(b) the vertical magnetic field $B_z$, 
					(c) the magnitude of magnetic shear $ \left|\nabla \times \mathbf{B}_h \right|$ of transverse magnetic field, 
					(d) gas pressure $P$, 
					(e) mass density $\rho$, 
					(f) vertical velocity $v_z$, 
					(g) kinetic energy density $Ek$, 
					and (h) magnetic energy density $E_{\mathrm{mag}}$. 
					Higher NMI values (shown in red) indicate stronger statistical dependence. 
					Blue, orange, and green stars denote the onset, decay, and dissipation 
					phases of the vortex as identified by the $\Gamma$-method.
				}
				
				\label{fig:MI_T_vortex}
			\end{figure*}

			To identify the key drivers of temperature evolution within the vortex, we computed the NMI between plasma temperature and a selection of physical variables linked to heating processes or thermodynamic coupling. Figure~\ref{fig:MI_T_vortex} presents the temporal and vertical variation of these dependencies within the vortex core.  The initial enhanced coupling between $v_z$, $\nabla \cdot \mathbf{v_h}$ and temperature is most pronounced during the vortex development and in the lower atmosphere, suggesting that the initial plasma cooling is driven by the jet transporting the vortex upwards and the initial flux tube expansion. As expected, plasma temperature initially shows strong statistical coupling with both density and pressure, consistent with the ideal gas law under quasi-static conditions. However, during the vortex phase, the coupling with plasma density significantly weakens above 1~Mm, indicating that the temperature evolution is no longer primarily governed by adiabatic compression or expansion. Instead, this indicates a breakdown of ideal thermodynamic behaviour due to non-adiabatic processes, such as current dissipation, wave heating, and magnetically induced flows. The reduced NMI between temperature and density/pressure reflects the growing influence of magnetic forces and energy injection mechanisms in the upper atmosphere.
			
			Although vortex-driven expansion initially leads to adiabatic cooling, this effect is short-lived. Within 10 time steps, the plasma temperature begins to rise, indicating that additional heating mechanisms rapidly become dominant. The sustained NMI between temperature and magnetic shear throughout the vortex lifetime suggests that current structures play a consistent role in modulating plasma heating. This coupling is particularly notable given that the magnetic shear-temperature association emerges primarily during the vortex phase, vanishing before and after. This timing supports the interpretation that currents, associated with twisted and evolving magnetic fields, contribute to localized heating within the vortex core.  The temporal alignment between this temperature rise and elevated magnetic shear points toward current dissipation as a likely source of energy input. Although this dissipation is numerical in origin, because of the lack of explicit resistivity in the simulation, it still reflects the physical tendency for currents to form and dissipate in regions of strong shear. 
			These results indicate that vortex-driven magnetic structuring channels energy into heat, particularly where plasma-$\beta$ is low (upper atmosphere)		
			
			\begin{figure*}
				\centering
				\includegraphics[width=\linewidth]{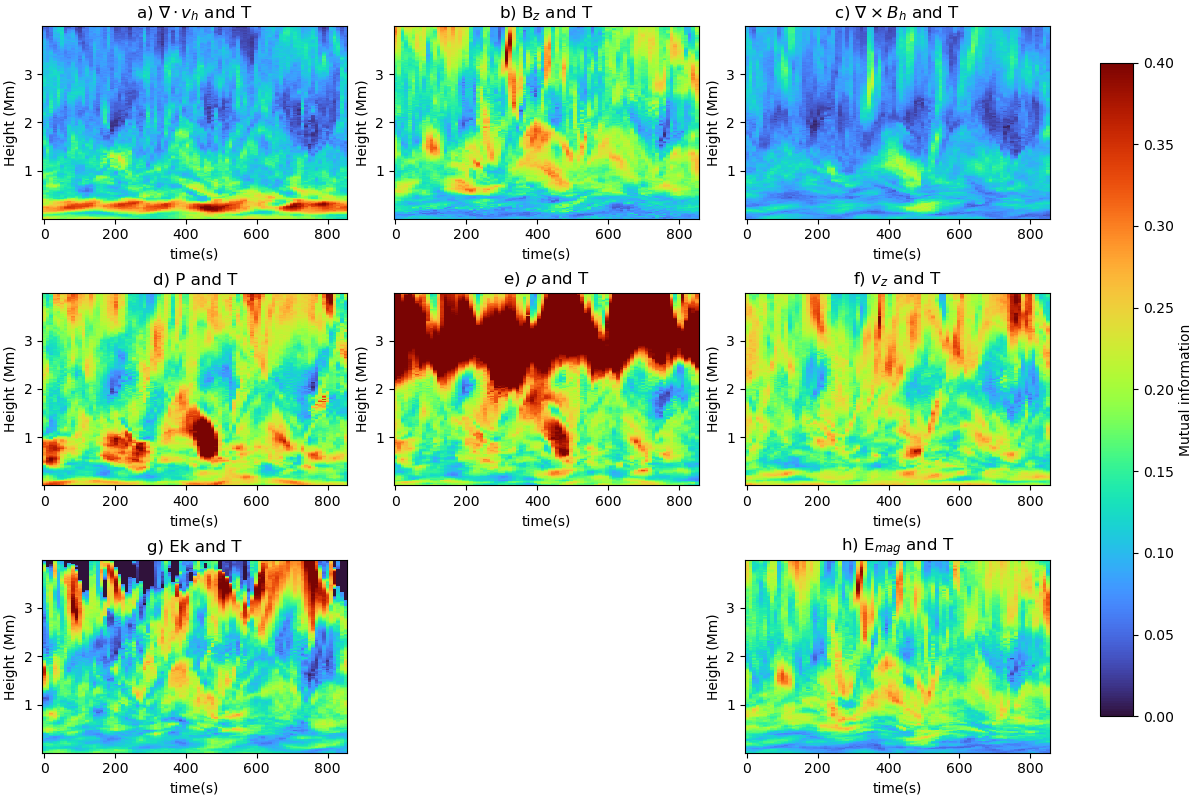}
				\caption{Same as Figure \ref{fig:MI_T_vortex}, but here we show the MI maps for the control region, i.e. a region where no coherent vortex is present, and the region contains only a slightly twisted magnetic flux tube.}
				\label{fig:MI_T_novortex}
			\end{figure*}

			\section{Discussion}
			To understand the vortex-driven dynamics in a simulated solar atmospheric plasma, we examined two large magnetic flux tubes: one that hosts a strong solar tornado with a coherent vortex flow and another characterized only by shear flows and transient swirls. The vortex flow first emerges in the upper photosphere and lower chromosphere. From there, it expands both upward and downward, forming a long-lived column of rotating plasma. It persists for approximately 5-10 minutes, with a horizontal radius of over 1~Mm in the upper atmosphere. Our analysis focuses on its evolution from the photosphere through the upper atmosphere, within the height range of 0-4~Mm. Throughout the full height of the vortex, the plasma-$\beta$ steadily decreases as the structure develops, reflecting a growing dominance of magnetic forces over gas pressure. Interestingly, the magnetic field intensifies at the vortex onset and remains strong even after the large-scale structure dissolves, pointing to lasting magnetic reconfiguration as reported by \cite{Diaz_2024}. Aside from a brief initial cooling, the vortex phase is marked by a persistent temperature increase, highlighting the efficiency of heating mechanisms operating throughout its lifetime.

			Using information-theoretic diagnostics, we quantified how vortex evolution alters the behaviour of key MHD variables. SE captures changes in the spatial complexity of individual fields, whereas NMI measures how relationships between different variables strengthen or weaken over time and height. This difference in what each metric probes is the sense in which they are complementary: SE describes the intrinsic structural evolution of a single variable, while NMI characterizes how that variable co-varies with others.
			The magnetic-field variables illustrate this complementarity clearly. The evolution of SE shows that the entropy of the vertical magnetic field increases during vortex onset in the upper atmosphere, indicating the development of stronger spatial gradients and more irregular $B_z$ distributions. As the vortex matures, $B_z$ entropy decreases, reflecting the emergence of a more ordered, collimated structure. In contrast, the magnetic-shear entropy starts low, rises sharply as currents intensify during vortex development, and falls again during the decay phase, consistent with current dissipation. Cross-comparison with NMI maps reveals that when magnetic-shear entropy peaks, temperature becomes strongly coupled to shear, indicating that thermal evolution is linked to the buildup and subsequent dissipation of current-related complexity.

			The thermodynamic variables also show clear stratification. In the lower atmosphere, SE indicates low entropy in density but high entropy in pressure, reflecting relatively uniform mass loading alongside strong compressive variability. NMI confirms that temperature remains persistently coupled to horizontal compression ($\nabla \cdot v_h$) throughout the vortex lifetime. At the same time, temperature also shows strong coupling with vorticity, which itself contributes to the velocity gradient. This dual dependence suggests that both compressive processes and viscous dissipation contribute to heating in the lower layers. In the upper atmosphere, the entropy patterns reverse: density entropy rises (stronger gradients and irregular distributions) while pressure entropy decreases (more ordered stratification). The coupling of temperature and density diminishes considerably in the upper part of the domain, indicating a departure from the ideal gas law. NMI reveals that temperature coupling with vorticity persists here, implying a continuing role for viscous heating, but now temperature also couples strongly to magnetic shear. This indicates that, higher up, magnetic structuring and current dissipation become additional drivers of thermal evolution alongside viscous effects. 
			These findings confirm that vortices intensify velocity and magnetic field gradients, establishing conditions conducive to energy dissipation, consistent with previous studies \citep{Yadav_2021, Silva_2024b, Silva_2024a}.
			This behaviour is consistent with previous studies \citep{Yadav_2021, Silva_2024b, Silva_2024a}.
			Our analysis extends this picture by showing that plasma temperature is statistically coupled to these gradients, demonstrating that the dissipated energy is effectively converted into heating within the vortex. This means that regions where the vortex generates strong shear or currents correspond to sites of effective energy release. 
			
			The coupling is not one-sided: plasma variables also influence vorticity. Elevated temperature can enhance pressure gradients, which in turn drive flows that sustain or modify vortical motions. In the lower atmosphere, where pressure entropy is high, pressure gradients are likely to reinforce rotational shear and help maintain vortex coherence. In the upper atmosphere, strong magnetic shear can exert Lorentz forces that realign flows, thereby shaping or even destabilizing the vortex. The observed NMI links, therefore, suggest a bidirectional relationship: vorticity drives gradients that heat the plasma, while thermal and magnetic structuring provide feedback that regulates the strength, persistence, and eventual decay of the vortex as reported by \citet{Shelyag_2012, Silva_2024b}.

			\section{Conclusions}
			
			We applied SE and NMI to a long-lived vortex in a Bifrost simulation to quantify how it reshapes its local environment. Our analysis leads to the following key conclusions:
			
			\begin{itemize}
				\item \textbf{The vortex transforms the local dynamical regime.} It suppresses the p-mode-like, pressure-dominated behaviour present in a neighbouring flux tube and replaces it with a magnetically controlled structure.
				
				\item \textbf{Magnetic complexity rises and falls with the vortex life cycle.} SE peaks during vortex growth, consistent with current formation and energy buildup, and decreases as the field becomes more ordered.
				
				\item \textbf{Heating pathways change with height.} Temperature responds mainly to compressional and expansion motions in the photosphere and lower chromosphere, while viscous and current-driven processes dominate in the upper atmosphere.
				
				\item \textbf{The plasma departs from adiabatic behaviour.} Temperature-density coupling weakens during the vortex phase, indicating that dissipative processes increasingly govern the thermal evolution.
				
				\item \textbf{The strong couplings are generated by the vortex itself.} A nearby non-vortical flux tube shows only weak and uniform NMI patterns, confirming that the enhanced links in the vortex region arise dynamically rather than from static co-location.
				
				\item \textbf{Information theory reveals interactions that traditional diagnostics miss.}  
				By distinguishing genuine dynamical coupling from mere spatial coincidence, NMI provides a quantitative view of how flows, fields, and thermodynamic variables interact during the vortex life cycle.
			\end{itemize}
			
			These results show that solar vortices are not passive features of turbulent flows. They actively reorganize the plasma, reshape magnetic fields, and drive localized, height-dependent heating.
			Information-theoretic tools also offer practical advantages for studying solar plasmas. NMI and SE provide a robust way to probe nonlinear interactions without relying on restrictive assumptions about the MHD equations. In the chromosphere, effects such as partial ionization, radiative losses, anisotropic conduction, and strong magnetic structuring limit the validity of the usual single-fluid, ideal-gas approach. This makes analytical separation of heating terms very difficult. The governing equations are highly nonlinear, coupled, and spread across multiple scales, which complicates direct term-by-term decomposition. In contrast, information-theoretic diagnostics quantify statistical dependencies directly from the data, allowing us to assess dynamical coupling and potential causal pathways even when the governing equations cannot be cleanly separated. As high-resolution chromospheric observations continue to improve, these methods offer a scalable, model-agnostic framework for interpreting complex magnetic structures such as vortices, swirls, and flux ropes, and they extend naturally to other turbulent and nonlinear astrophysical systems.

			\section*{Acknowledgments}
			S.S.A.S. and V.F.  are grateful to the Science and Technology Facilities Council (STFC) grants ST/V000977/1, ST/Y001532/1. V.F., S.S.A.S., I.B. and G.V. are thankful to The Royal Society, International Exchanges Scheme, collaboration with Greece (IES/R1/221095) and The Royal Society, IEC/R3/233017 - International Exchanges 2023 Cost Share (NSTC), collaboration with Taiwan. E.L.R. acknowledges the Brazilian agency CNPq under grant 306920/2020-4. S.S.A.S. and V.F. would like to thank the International Space Science Institute (ISSI) in Bern, Switzerland, for the hospitality provided to the members of the team on `Opening new avenues in identifying coherent structures and transport barriers in the magnetised solar plasma’.  S.S.A.S. is grateful to Iñaki Esnaola for insightful discussions that contributed to this work.

			\bibliographystyle{aasjournal}
			\bibliography{biblio}
		\end{document}